\documentclass{article}
\usepackage[margin=1in]{geometry}

\usepackage{amsmath,amssymb}
\usepackage{graphicx}
\usepackage[utf8]{inputenc}
\usepackage{textcomp}
\usepackage{cite}
\usepackage{nameref,hyperref}
\usepackage{array}
\usepackage{booktabs}
\usepackage[frozencache,cachedir=minted-cache]{minted}
\usepackage{multirow}
\usepackage{setspace}
\usepackage[T1]{fontenc}
\usepackage{standalone}
\usepackage{tikz}
\usepackage{caption}
\usepackage{subcaption}
\usetikzlibrary{fit, positioning, arrows.meta, calc, backgrounds}
\definecolor{apibg}{RGB}{245,245,235}
\definecolor{uibg}{RGB}{220,235,250}

\setminted{baselinestretch=0.9}

\title{TreeFlow: Probabilistic Modelling and Automatic Differentiation for Phylogenetics
}
\author{
Christiaan Swanepoel,$^{\ast, 1, 2}$\\
Mathieu Fourment,$^{3}$\\
Xiang Ji,$^{4}$\\
Hassan Nasif,$^{5}$\\
Marc A Suchard,$^{6,7,8}$\\
Frederick A Matsen IV,$^{5,9,10}$\\
Alexei J Drummond$^{1,11}$
}

\date{}

\begin{document}
\maketitle

\noindent {\small
    ${^1}$Centre for Computational Evolution, The University of Auckland, Auckland, New Zealand;\\
$^{2}$School of Computer Science, The University of Auckland, Auckland, New Zealand, 1010;\\
$^{3}$Australian Institute for Microbiology and Infection, University of Technology Sydney, Ultimo NSW, Australia;\\
$^{4}$Department of Mathematics, Tulane University, New Orleans, USA;\\
$^{5}$Public Health Sciences Division, Fred Hutchinson Cancer Research Center, Seattle, Washington, USA;\\
$^{6}$Department of Human Genetics, University of California, Los Angeles, USA;\\
$^{7}$Department of Computational Medicine, University of California, Los Angeles, USA;\\
$^{8}$Department of Biostatistics, University of California, Los Angeles, USA;\\
$^{9}$Department of Statistics, University of Washington, Seattle, USA;\\
$^{10}$Department of Genome Sciences, University of Washington, Seattle, USA;\\
$^{11}$School of Biological Sciences, The University of Auckland, Auckland, New Zealand, 1010;\\
}

\bigskip

\noindent {\small\itshape Correspondence to be sent to: School of Computer Science, The University of Auckland, Auckland, New Zealand, 1010;\\
Email: cswa648@aucklanduni.ac.nz
}

\bigskip

\section*{Abstract}
Probabilistic modelling frameworks are powerful tools for statistical modelling and inference. They are not immediately generalizable to phylogenetic problems due to the particular computational properties of the phylogenetic tree object. TreeFlow is a software library for probabilistic modelling and automatic differentiation with phylogenetic trees. It embeds phylogenetic trees in the TensorFlow Probability framework, and implements inference algorithms for phylogenetic models given a fixed tree topology. We demonstrate how TreeFlow can be used to quickly implement and assess new models. We also show that it provides reasonable performance for gradient-based inference algorithms compared to specialized computational libraries for phylogenetics.

\medskip
\noindent\textbf{Keywords: }phylogenetics; Bayesian inference; probabilistic modelling; variational inference

\section{Introduction}

Traditionally, Bayesian phylogenetic analyses have been performed by specialized software \cite{huelsenbeck2001mrbayes, lartillot2009phylobayes, drummond2012bayesian}. A number of software packages exist that implement a broad but predefined collection of models and associated specialized inference methods. Typically, inference is handled by carefully crafted but computationally costly Markov chain Monte Carlo (MCMC) methods \cite{metropolis1953equation, hastings1970monte}. Considerable effort has been invested in improving the computational efficiency of MCMC for phylogenetics \cite{mateiu2006inferring, lakner2008efficiency, hohna2012guided, zhang2020improving}. In contrast, in other realms of statistical analysis, \textit{probabilistic graphical modelling} or \textit{probabilistic programming} software libraries have entered into widespread use. These allow the specification of almost any model as a probabilistic program, and inference is provided automatically with a generic inference method. Exploiting the power of these tools in phylogenetic analyses could significantly accelerate research by making the process of developing new models and implementing inference faster and more flexible, as evidenced by recent work attempting to reconcile probabilistic modelling and programming frameworks with phylogenetic models \cite{fourment2019evaluating, ronquist2021universal}.

Probabilistic modelling tools, such as BUGS \cite{lunn2000winbugs}, Stan \cite{carpenter2017stan}, PyMC3 \cite{salvatier2016probabilistic}, and TensorFlow Probability \cite{dillon2017tensorflow} allow users to specify probabilistic models by describing the structure of a \textit{probabilistic graphical model} with code. Typically, this involves composing pre-implemented probability distributions by specifying conditional dependencies between variables. This structure can be used to evaluate a joint probability density function for the model variables. Advancements in automatic inference methods have allowed these tools to perform efficient inference on a wide range of models. Some notable examples, including automatic differentiation variational inference \cite{kucukelbir2017automatic} (ADVI) and Hamiltonian Monte Carlo \cite{duane1987hybrid} (HMC), use local gradient information from the model's probability density function to efficiently navigate the parameter space.

Another class of methods for performing statistical inference are \textit{probabilistic programming languages}. These provide more flexibility than graphical modelling methods by explicitly describing the generative process of the model as a \textit{probabilistic program}, rather than just composing probability distributions, and typically include inference schemes that do not require explicit evaluation of a joint probability density. Since the term \textit{probabilistic programming} has been widely adopted to describe probabilistic graphical modelling software, probabilistic programming languages are often qualified as \textit{universal} in that they are not restricted to composing pre-implemented probability distributions. Examples of universal probabilistic programming languages include Church \cite{goodman2008church} and Pyro \cite{bingham2019pyro}.

One key technology that enables gradient-based automatic inference with probabilistic models is \textit{automatic differentiation}. Automatic differentiation refers to methods which efficiently calculate machine-precision gradients of functions, specified by computer programs, without extra analytical derivation or excessive computational overhead compared to the function evaluation. Numerical programs consist of elementary arithmetic operations, and automatic differentiation tools apply the chain rule to these to compute derivatives directly, without requiring the construction of explicit mathematical expressions. Automatic differentiation frameworks that have statistical inference packages built on top of them include Theano \cite{bergstra2010theano}, Pytorch \cite{paszke2019pytorch}, JAX \cite{jax2018github} and TensorFlow \cite{abadi2016tensorflow}. Some of these extend to non-trivial computational constructs such as control flow and recursion \cite{yu2018dynamic}.

The structure of the phylogenetic tree object is a major barrier to implementing efficient inference for phylogenetic probabilistic models. It is not clear how the association between its discrete and continuous quantities (the topology and branch lengths respectively) should be represented and handled in inference. Also, the combinatorial explosion of the size of the discrete part of the phylogenetic state space presents a major challenge to any inference algorithm. Generic random search methods for discrete variables, as in the naive implementation of MCMC sampling, do not scale appropriately to allow inference on modern datasets with thousands of taxa.

Efforts have already been made to apply probabilistic graphical modelling and probabilistic programming to phylogenetic methods
\cite{hohna2014probabilistic, fourment2019evaluating, ronquist2021universal, drummond2023linguaphylo}. RevBayes \cite{hohna2016revbayes} implements a probabilistic modelling language that is focused on phylogenetic models. However, it requires the user to manually specify an MCMC scheme for performing inference. Universal probabilistic programming languages have also been used to express generative processes for trees and to automatically generate Sequential Monte Carlo inference schemes for complex speciation models. It has been proposed that the same languages could generate inference schemes for other phylogenetic models, including MCMC samplers over tree topologies \cite{ronquist2021universal, senderov2024treeppl}. Blang \cite{bouchard2022blang} is a probabilistic modelling framework which supports arbitrary data structures that has been used to implement phylogenetic models and can automatically construct inference schemes. Another tool, LinguaPhylo, provides a domain specific graphical modelling language for phylogenetics and has the capability to automatically generate a specification for an MCMC sampler for inference \cite{drummond2023linguaphylo}.

Applying gradient-based inference methods to phylogenetics is a major challenge. \textit{Probabilistic path Hamiltonian Monte Carlo} has been developed to use gradient information to sample phylogenetic posteriors across multiple tree topologies \cite{dinh2017probabilistic}, though moves between tree topologies are not totally informed by gradient information and are similar to the random-walk proposal distributions available to standard MCMC routines. Hamiltonian Monte Carlo proposals for continuous parameters within tree topologies has been paired with standard MCMC moves between topologies for estimating local mutation rates and divergence times using scalable algorithms for gradient calculation \cite{fisher2023shrinkage, ji2023scalable}. Another approach has used Stan to apply automatic differentiation variational inference to phylogenetic inference with a fixed rooted tree topology \cite{fourment2019evaluating}. Variational inference has been applied to phylogenetic tree inference using a clade-based distribution on unrooted tree topologies \cite{zhang2018variational}. This approach has been extended to a more expressive family of approximating distributions \cite{zhang2020improved}, and applied to rooted phylogenies \cite{zhang2022variational}. Not all variational approaches to phylogenetics are gradient-based: VaiPhy performs approximate inference in an augmented tree space using coordinate ascent update equations, together with a proposal distribution for tree topologies and a scheme for sampling branch lengths directly under the Jukes-Cantor model, so it does not require automatic differentiation \cite{koptagel2022vaiphy}.

TreeFlow aims to enable the application of gradient-based inference methods to phylogenetic models. It implements a representation of phylogenetic trees in the TensorFlow computational framework which supports automatic differentiation, and a range of associated computations typically used in phylogenetic models of sequence evolution. These are all integrated with the TensorFlow Probability probabilistic modelling framework. TreeFlow provides utilities for applying variational Bayesian inference to phylogenetic models, particularly with a fixed tree topology, and a command line user interface.

\section{Software Description}

TreeFlow is a Python library which provides modular tools for phylogenetic modelling and inference. It provides two primary interfaces: a user interface for performing parameter inference in fixed-topology models of a standard form, and a Python developer API for implementing and composing custom phylogenetic models. TreeFlow supports both rooted and unrooted phylogenetic trees, but the user interface is focused on inference with rooted trees. The architecture of the TreeFlow package is displayed in Figure \ref{fig:architecture}.

\newlength{\originalbaselineskip}
\setlength{\originalbaselineskip}{\baselineskip}

\begin{figure}
    \centering
\begin{tikzpicture}[
        x=1cm, y=1cm,
        box/.style={draw, rounded corners=2pt, fill=white, align=center,
                    text width=4.1cm, inner xsep=0.14cm, inner ysep=0.12cm,
                    minimum height=1.05cm, font=\fontsize{7}{8.4}\selectfont},
        depbox/.style={box, rounded corners=0pt, fill=black!6},
        apifitbox/.style={draw, thick, dashed, rounded corners=4pt,
                          inner sep=0.3cm, fill=apibg},
        uifitbox/.style={draw, thick, dashed, rounded corners=4pt,
                         inner sep=0.3cm, fill=uibg},
        grouplabel/.style={font=\bfseries\fontsize{8}{9.6}\selectfont,
                           inner sep=0pt, anchor=north west},
        arrow/.style={-{Stealth[length=4.5pt,width=3.2pt]}, semithick, black!75},
        bus/.style={semithick, black!75},
        junction/.style={circle, fill=black!75, inner sep=0pt, minimum size=2.6pt},
    ]

    \renewcommand{\baselinestretch}{1}\selectfont

    \def\dsc{\fontsize{6}{7}\selectfont\color{black!55}}

    \node[box] (tree)          at ( 5.0,  0.0) {\texttt{treeflow.tree}\\{\dsc Tree representation}};
    \node[box] (native)        at (10.0,  0.0) {\texttt{treeflow.acceleration.native}\\{\dsc Native likelihood and ratio transform ops}};

    \node[box] (io)            at ( 5.0, -1.45) {\texttt{treeflow.tree.io}\\{\dsc Tree reading and writing}};
    \node[box] (traversal)     at (10.0, -1.45) {\texttt{treeflow.traversal}\\{\dsc Tree traversals}};

    \node[box] (priors)        at ( 0.0, -2.9) {\texttt{treeflow.distributions.tree}\\{\dsc Tree priors}};
    \node[box] (approximation) at ( 5.0, -2.9) {\texttt{treeflow.model.approximation}\\{\dsc Tree posterior approximations}};
    \node[box] (bijectors)     at (10.0, -2.9) {\texttt{treeflow.bijectors}\\{\dsc Bijections on times on trees}};

    \node[box] (evolution)     at ( 0.0, -4.35) {\texttt{treeflow.evolution}\\{\dsc Substitution models}};
    \node[box] (vi)            at ( 5.0, -4.35) {\texttt{treeflow.vi}\\{\dsc Fixed topology variational inference}};

    \node[grouplabel] (apilabel) at (-2.19, 0.525) {Developer API};

    \begin{pgfonlayer}{background}
        \node[apifitbox, fit=(apilabel)(tree)(native)(io)(traversal)
                             (priors)(approximation)(bijectors)(evolution)(vi)]
              (developerapi) {};
    \end{pgfonlayer}

    \node[box] (phylomodel) at (2.5, -6.7) {\texttt{treeflow.model.phylo\_model}\\{\dsc Model specification}};
    \node[box] (cli)        at (7.5, -6.7) {\texttt{treeflow.cli}\\{\dsc Inference command line interfaces}};

    \node[grouplabel] (uilabel) at (-2.19, -6.175) {User interface};
    \coordinate (uipad) at (12.19, -6.7);

    \begin{pgfonlayer}{background}
        \node[uifitbox, fit=(uilabel)(phylomodel)(cli)(uipad)] (userinterface) {};
    \end{pgfonlayer}

    \node[depbox] (tfp)         at ( 0.0, 1.65) {\texttt{tensorflow\_probability.}\\\texttt{distributions}};
    \node[depbox] (tensorarray) at (10.0, 1.65) {\texttt{tensorflow.TensorArray}};

    \draw[arrow] (tree) -- (io);
    \draw[arrow] (tree.south east) -- (traversal.north west);
    \draw[arrow] (tree.west) -| (priors.north);
    \draw[arrow] (native) -- (traversal);
    \draw[arrow] (traversal) -- (bijectors);
    \draw[arrow] (bijectors.west) -- (approximation.east);
    \draw[arrow] (approximation) -- (vi);

    \draw[arrow] (priors.east) -- (2.5,-2.9) -- (phylomodel.north);
    \draw[arrow] (evolution.south) -- ([xshift=-1.2cm]phylomodel.north);
    \draw[arrow] (io.east) -- (7.5,-1.45) -- (cli.north);
    \draw[arrow] (vi.south) -- ([xshift=-1.2cm]cli.north);
    \draw[arrow] (phylomodel) -- (cli);

    \draw[bus]   (tfp.west) -- (-2.89,1.65) -- (-2.89,-6.7);
    \draw[arrow] (-2.89,-2.9) -- (priors.west);
    \draw[arrow] (-2.89,-4.35) -- (evolution.west);
    \draw[arrow] (-2.89,-6.7) -- (phylomodel.west);
    \node[junction] at (-2.89,-2.9) {};
    \node[junction] at (-2.89,-4.35) {};

    \draw[arrow] (tensorarray.east) -- (12.89,1.65) -- (12.89,-1.45) -- (traversal.east);

    \path (-3.05,0) (13.05,0);

\end{tikzpicture}
    \caption{Diagram showing the architecture of the TreeFlow package. Nodes represent Python modules and the arrows between them denote major dependencies. The dashed rectangles separate the components based on their primary interfaces. Nodes outside the two interface groups denote major external dependencies.}
    \label{fig:architecture}
\end{figure}

\setlength{\baselineskip}{\originalbaselineskip}

We first describe the user interface, which provides the most accessible entry point for users wanting to apply TreeFlow to standard phylogenetic analyses. We then describe the developer API, which provides the modular components that underpin the user interface and allow users to implement new models.

\subsection{User interface}

TreeFlow provides command line interfaces for fixed-topology inference. These allow inference on standard phylogenetic models such as those performed by specialized software. For inputs, they take a nucleotide sequence alignment in the FASTA or Nexus format, a rooted tree topology in Newick format, and a TreeFlow model specification in YAML format. Command line interfaces are provided for both automatic differentiation variational inference and maximum a posteriori inference.

Because many phylogenetic analyses use models of a similar structure, TreeFlow provides a format for concisely specifying this kind of model. A single unpartitioned multiple sequence alignment can be modelled with a single substitution model, tree prior, and associated parameter priors. The model specification is a nested dictionary, with keys at various levels corresponding to model components, their parameters, and distributions used as priors. This is transformed into a TensorFlow Probability joint distribution. The model definition structure can be serialized in the YAML markup format which provides a text-based model definition format. An example of this is shown in Supplementary Appendix Figure S1. The substitution and tree models can be drawn from the models described below, as well as parameter priors from a range of standard probability distributions implemented in TensorFlow Probability. This model definition format is not intended to cover the full range of probabilistic models that can be implemented with TreeFlow's Python API, but rather provide a convenient input for TreeFlow's command line interfaces. Installation instructions, a complete reference for the YAML model definition format and its available model components, and a step-by-step tutorial are provided in the online manual (\url{https://treeflow.readthedocs.io}). New users may find the ``rates and dates'' tutorial, which works through a full analysis using this format, a helpful starting point after installing the software.

\subsection{Developer API}

The developer API allows users to implement new models and compose them with existing model components in a flexible structure, while leveraging the inferential machinery provided by TreeFlow. The TreeFlow Python library includes modules for the representation, traversal, and reading and writing of phylogenetic trees. There are modules implementing bijections for phylogenetic trees, phylogenetic posterior approximations, and fixed-topology variational Bayesian inference. It also implements tree priors and models of sequence evolution. TensorFlow Probability distributions can be composed into a \textit{probabilistic graphical model} using TensorFlow Probability's joint distribution functionality \cite{piponi2020joint}. The code to specify a joint distribution provides a concise representation of the model used in a data analysis. The ability to implement phylogenetic models in this framework means that automatic inference algorithms implemented in TensorFlow can be leveraged, as well as other utilities such as sampling from the joint prior. The subsections below describe the major components of the developer API.

\subsection{Tree representation}

TreeFlow is built on TensorFlow, a computational framework for machine learning. TreeFlow leverages TensorFlow's capabilities for accelerated numerical computation and automatic differentiation. TensorFlow operations can be executed on either the CPU or GPU, and as a result TreeFlow can make use of either type of processor. Many of TensorFlow's CPU operations are parallelized, allowing TreeFlow to make use of multiple processor cores. All of the benchmarks and examples in this paper, and TreeFlow's command line interfaces, use the CPU for computation.

TensorFlow's core computational object is the Tensor, a multi-dimensional array with a uniform data type. Phylogenetic trees are not immediately at home in a Tensor-centric universe, as they are a complex data structure, often defined recursively, with both continuous and discrete components. TreeFlow represents trees as a structure of Tensors. This structure contains floating point Tensors representing the times and branch lengths, and integer Tensors representing the topology, all encapsulated in a Python object. The topological Tensors include indices of node parents, children, and for pre- and post-order traversals. TensorFlow has extensive support for structures of Tensors such as the TreeFlow tree class, including using them as arguments to compiled functions and defining probability distributions over complex objects. This means computations and models involving phylogenetic trees can be expressed naturally.

\subsection{Tree distributions}
\label{treepriors}

TreeFlow uses the existing probabilistic modelling infrastructure provided by TensorFlow Probability, which implements standard statistical distributions and inferential machinery \cite{dillon2017tensorflow}. The Distribution interface encapsulates functionality for sampling from a distribution and computing its probability density or mass function. TreeFlow implements a tree distribution base class which allows TensorFlow Probability Distributions to be defined on TreeFlow's phylogenetic tree representation.

TreeFlow implements Distributions for some commonly-used generative models for phylogenetic trees.
The models currently implemented, Kingman's coalescent \cite{kuhner1995estimating} and the Birth-Death-Sampling speciation process \cite{stadler2009incomplete}, can be used as \textit{tree priors} in phylogenetic probabilistic models and to infer parameters related to population dynamics from genetic sequence data. Since sampling from these distributions is non-trivial and not required for variational inference, only the probability densities of these models are currently implemented.

\subsection{Models of sequence evolution}
\label{evolution}

The models of nucleotide sequence evolution currently implemented in TreeFlow are the Jukes-Cantor \cite{jukes1969evolution}, HKY85 \cite{hasegawa1985dating}, and General Time Reversible (GTR) \cite{tavare1986some} models. TreeFlow includes a standard approach for dealing with heterogeneity in mutation rate across sites based on marginalizing over a discretized site rate distribution \cite{yang1994maximum}. The probabilistic modelling framework, however, allows for the use of any appropriate distribution as the base site rate distribution rather than just the standard single-parameter Gamma distribution. For example, it is straightforward to replace the base Gamma distribution with a Weibull distribution, which has a quantile function that is much easier to compute \cite{fourment2019evaluating}. Thanks to TensorFlow's vectorized arithmetic operations, it is also natural to model variations in mutation rate over lineages by specifying parameters for multiple rates (possibly with a hierarchical prior) and multiplying by the branch lengths of the phylogenetic tree. Models which can be naturally expressed this way include the log-Normal random local clock \cite{drummond2006relaxed}, which is implemented with TreeFlow's model definition format, and auto-correlated relaxed clock models \cite{thorne1998estimating}, which would be straightforward to implement with the traversal module in the Python API.

\subsection{Tree traversals}

To perform model computations involving the phylogenetic tree, it is necessary to traverse the tree according to the structure specified by its topology. Since this structure is less accessible in an array-based tree representation, and native Python control flow cannot be used when the topology of the tree is a dynamic variable, TreeFlow provides utilities to perform tree traversals. These are used to efficiently implement some core computations and also made accessible to developers for extending TreeFlow's functionality.

When iterating over the nodes of the tree in an automatic differentiation framework it is important to consider computational complexity. Since modern genomic datasets can result in the number of nodes in the tree being extremely large, it is crucial to maintain linear scaling with the size of the tree. Any computation that produces values for every node and requires access to results for multiple previous nodes in the traversal might incur a quadratic computational cost. This could occur when the topology is a dynamic variable as, at a given step of the iteration, values from any element of the Tensor representation might need to be accessed. Since TensorFlow's automatic differentiation framework depends on the immutability of Tensors, an intermediate representation of the state for the whole tree needs to be maintained for every step of the iteration, leading to a quadratic complexity. Intuitively, because a Tensor cannot be modified in place, a step of the traversal that updates the value at one node cannot simply overwrite that entry; it must instead produce a whole new copy of the tree-wide state Tensor. Repeating this for every node produces as many copies of the state as there are nodes, each of that same size, all of which must be retained and differentiated through, so both the computation and the memory grow with the square of the number of nodes.

TreeFlow's solution to this scaling issue is to implement tree traversals using TensorFlow's \texttt{TensorArray} construct \cite{yu2018dynamic}. The TensorArray is a data structure representing a collection of tensors. The TensorArray relaxes the requirement of immutability to a \textit{write-once} property enforced on its constituent tensors. The TensorArray tracks dependencies between values so automatic differentiation can appropriately propagate gradients back through computation in linear time.

A notable computation that makes use of the tree traversal framework described above is the \textit{phylogenetic likelihood}, the probability of a sequence alignment given a phylogeny and model of sequence evolution. This typically involves integrating out character states at unsampled ancestral nodes using a \textit{dynamic programming} computation known as \textit{Felsenstein's pruning algorithm} \cite{felsenstein1981evolutionary}. Considerable attention has been given to the problem of efficiently computing gradients of branch lengths with respect to the phylogenetic likelihood \cite{ji2020gradients}. TreeFlow implements the pruning algorithm as a TensorArray-based \textit{postorder} traversal. The recursion is expressed in terms of the per-branch transition probability matrices rather than directly in terms of branch lengths and substitution model parameters. This is a lower-level interface for gradient calculation than that of specialized library BEAGLE \cite{ayres2019beagle}, which take branch lengths and substitution parameters as inputs and return the likelihood together with analytically derived gradients. Working at the level of transition probabilities is a more natural fit for automatic differentiation, because the gradient of the likelihood propagates back through the transition-probability computation to whatever parameters produced it. Consequently the branch-length gradient, and gradients with respect to any substitution model parameters, are obtained automatically without a hand-derived gradient expression for each model, and the branch-length gradient demonstrates linear scaling with respect to the number of taxa in our benchmarks (see the Benchmarks section below).

Because this pure-TensorFlow traversal obtains its gradients automatically and efficiently, it makes implementing a new model straightforward: a developer can compose the traversal with new model components and rely on automatic differentiation for the gradients, without writing native C++ code or deriving gradient expressions by hand. Where maximum performance is required, TreeFlow additionally provides a compiled C++ implementation of the pruning likelihood as a TensorFlow custom operation. This native operation performs the same recursion in native code, parallelized across alignment sites, and is paired with a hand-derived analytic gradient that reuses the partial likelihoods computed during the forward pass. It is registered with TensorFlow's automatic differentiation and selected by a single flag, so it is a drop-in accelerated alternative to the TensorArray traversal; its only additional requirement is a compilation step for the target platform.

Another useful tool for phylogenetic inference implemented in TreeFlow is the \textit{node height ratio transform} \cite{kishino2001performance}. This has been used to infer times on phylogenetic trees using maximum likelihood in the PAML software package \cite{yang2007paml}. The ratio transform parametrizes the internal node heights of a tree as the ratio between a node's height and its parent's height. This has been applied to Bayesian phylogenetic inference in the context of automatic differentiation variational inference \cite{fourment2019evaluating} and Hamiltonian Monte Carlo \cite{ji2023scalable}. The computation of node heights from ratios is implemented in TreeFlow as a TensorArray-based \textit{preorder} traversal. As with the phylogenetic likelihood, this transform is also provided as an optional compiled C++ TensorFlow custom operation, paired with a hand-derived analytic gradient, for use when performance is critical.

\subsection{Bijectors on phylogenetic trees}

The ratio transform described above is wrapped into a TensorFlow Probability Bijector which provides an interface for transforming real-valued distributions into phylogenetic tree distributions. The Bijector is an interface which represents a transformation of samples from a multivariate probability distribution. It requires implementation of the forward transformation and its inverse (which necessarily exists if the transformation is a bijection), as well as the determinant of the Jacobian matrix of the transformation. This can be used by TensorFlow Probability to perform change of variable, rewriting the probability density of the base distribution in the space the bijector transforms it to.

The ratio transformation has a triangular Jacobian matrix, which means computing the determinant required for the change of variable formula can be computed in linear time with respect to the number of internal node heights \cite{fourment2019evaluating}. Using change of variable, a multivariate distribution on ratios can be transformed into a distribution on the internal node heights of a rooted phylogeny. Other TensorFlow Probability Bijectors can be used to transform real-valued distributions into ratios. Another Bijector implemented in TreeFlow combines node heights with a fixed tree topology to produce a complete phylogenetic tree object. These bijectors can be composed sequentially, for example transforming unconstrained real values to ratios, then to node heights, then to a full tree, yielding a probability distribution over branch lengths for a given topology. The result can be used as an approximation to a phylogenetic posterior distribution and combined with models such as TreeFlow's tree priors and phylogenetic likelihood.

\subsection{Variational inference}

One form of statistical inference for which gradient computation is essential is variational Bayesian inference \cite{jordan1999introduction}. The goal of Bayesian inference is to characterise a \textit{posterior distribution} which represents uncertainty over model parameters. Variational Bayesian inference achieves this by optimizing an approximation to the posterior distribution which has more convenient analytical properties. The approximation is fitted by minimizing the Kullback-Leibler divergence from the approximation to the posterior, which is equivalent to maximizing a lower bound on the marginal likelihood known as the \textit{evidence lower bound} (ELBO). This direction of the divergence penalizes putting mass where the posterior has none much more heavily than the reverse, so a fitted approximation tends to settle on a region of high posterior density and to understate posterior variance. Typically, the optimization routine used in variational inference can scale to a larger number of model parameters than the random-walk sampling methods used by MCMC methods.

One concrete implementation of variational inference is \textit{automatic differentiation variational inference} (ADVI) \cite{kucukelbir2017automatic}. ADVI can perform inference on a wide range of probabilistic graphical models composed of continuous variables. It automatically constructs an approximation to the posterior by transforming a convenient base distribution to respect the domains of the model's component distributions. It then optimizes this approximation using stochastic gradient methods \cite{robbins1951stochastic, bottou2010large}. Typically, independent Normal distributions are used as the base distribution for computational convenience. This is known as the \textit{mean field approximation}, and for posterior distributions that have significant correlation-structure, skew, multi-modality, or heavy tails, can introduce error in parameter estimation \cite{blei2017variational}. The standard remedy for correlation structure is the \textit{full rank approximation}, which gives the base distribution a full covariance matrix so that linear correlation between every pair of coordinates can be represented, at the cost of a number of variational parameters quadratic in the dimension of the model. More generally, possible solutions to this approximation error include highly flexible variational approximations with large numbers of parameters \cite{rezende2015variational} or structured approximations that are informed by the form of the model \cite{ambrogioni2021automatic}.

TreeFlow leverages the stochastic gradient optimizers already implemented in TensorFlow and the variational inference objectives in TensorFlow Probability. The discrete topology element of phylogenetic trees is an obstacle in the usage of these algorithms, which are typically restricted to continuous latent variables. Often, the phylogenetic tree topology is not the latent variable of interest, and is not a significant source of uncertainty \cite{yang2000codon}. This can be the case when divergence times or other substitution or speciation model parameters are the focus. In this scenario, useful results can be obtained by performing inference with a fixed tree topology, such as one obtained from fast maximum likelihood methods. This is the approach taken by the NextStrain platform \cite{hadfield2018nextstrain}, which uses the scalability afforded by a fixed tree topology to allow large-scale rapid phylodynamic analysis of pathogen molecular sequence data. Using a variational approximation with a fixed topology in TreeFlow makes the application of ADVI to phylogenetic models straightforward. When using variational inference in TreeFlow, care must be taken if it is possible the tree topology for a given dataset is difficult to infer by assessing how sensitive the inference results are to it.

Although TreeFlow's current variational inference scheme is limited to a fixed tree topology, the rest of the library is not constrained in this manner. The Tensor-based tree representation means that the topology is part of TensorFlow's computational graph. Additionally, computations such as the phylogenetic likelihood are implemented with support for a dynamic tree topology. This means that existing approaches to variational inference over tree topologies \cite{zhang2018variational} could be implemented using TreeFlow's model specification and likelihood evaluation, extending the library to support topology inference in the future.

\subsection{Model approximations}

TreeFlow implements posterior approximations for ADVI using TensorFlow Probability's bijector framework to transform a base distribution. Approximations are constructed automatically from probabilistic models in the form of TensorFlow Probability joint distribution objects. A base distribution, by default an uncorrelated Normal, is split into the variables of the joint distribution, and transformed to match the support of the model variables. The base distribution for the phylogenetic tree is transformed using a chain of bijectors. Learnable parameters to optimize are introduced by adding an initial trainable transformation to the base distribution. For the mean field approximation, this is an affine elementwise shifting and scaling operation.

TreeFlow implements both the mean field and full rank approximations described above. Neither is a good fit for a fixed-topology phylogenetic model. The mean field approximation cannot represent posterior correlation at all, and the correlations it discards are the ones that matter most: the sequence data constrain the product of the evolutionary rate and the age of the tree far better than either quantity separately, so the clock rate, the tree prior's rate parameter and the root height are strongly correlated in the posterior. The full rank approximation can represent those correlations, but its number of variational parameters is quadratic in the number of taxa and is dominated by the covariances between node heights, which is a large cost for structure concentrated in a handful of coordinates.

TreeFlow therefore provides a third approximation, which we call \textit{root full rank}. It places a full covariance block over the model parameters together with the root height of the tree, and treats the remaining node-height ratios as independent. The number of variational parameters is then linear in the number of taxa, as for the mean field approximation, while the correlations between the overall scale of the tree and the model parameters are represented as they are under the full rank approximation. Models may additionally contain parameters that are replicated once per branch, whose dimension grows with the tree in the same way the node heights do. An example is the per-lineage transition-transversion ratio model of the carnivores analysis below. These are excluded from the covariance block by naming them when the approximation is constructed, and are approximated independently alongside the node-height ratios. The root full rank approximation is correspondingly less well suited to models where branch-dependent correlation structure exists, which we return to in the Discussion. The three approximations are compared on both of the datasets analyzed below in the Supplementary Appendix (Figures S3 and S4).

ADVI opens the door to using TensorFlow's neural network framework to implement deep-learning-based variants such as variational inference with \textit{normalizing flows} \cite{rezende2015variational}, which transform the base distribution through invertible trainable neural network layers to better approximate complex posterior distributions. Approximations based on \textit{inverse autoregressive flows} \cite{kingma2016improved} are implemented in TreeFlow, but since their computational and memory requirements scale quadratically with the number of model variables they are not ideal for large phylogenetic analyses.

\section{Biological examples}

We used TreeFlow to perform fixed-topology phylogenetic analyses of two datasets. The first is an alignment of 62 mitochondrial DNA sequences from carnivores \cite{suchard2009many}. The second is a dataset of 980 influenza sequences \cite{vaughan2014efficient}. In both datasets maximum likelihood unrooted topologies are estimated using RAxML \cite{stamatakis2014raxml}. These topologies are rooted with Least Squares Dating \cite{to2016lsd}.

\subsection{Probabilistic modelling and model selection}

In the carnivores dataset, we demonstrate the flexibility of probabilistic modelling with TreeFlow by investigating variation in the ratio of transitions to transversions in the nucleotide substitution process across lineages in the tree. Early maximum likelihood analyses of mitochondrial DNA in primates detected variation in this ratio, but without a biological basis, it was attributed to a saturation effect \cite{yang1999estimation}. A later simulation-based investigation showed that this was a reasonable explanation, which exposed the limitations of nucleotide substitution models for estimating the length of branches deep in the tree \cite{duchene2015declining}.

\begin{figure}
    \centering
    \includegraphics[width=\linewidth]{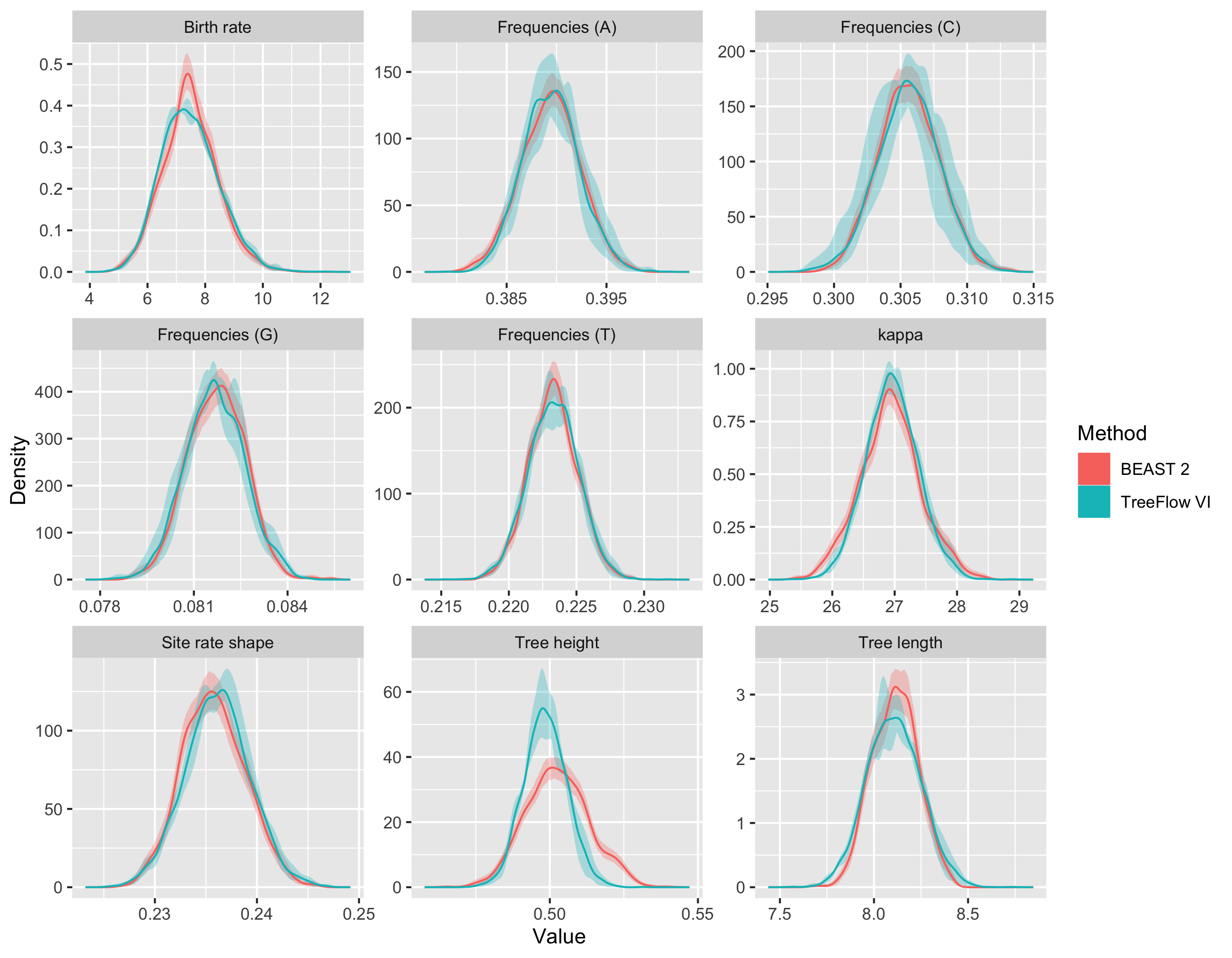}
    \caption{Posterior marginal parameter estimates for the carnivores base model (a single transition/transversion ratio, kappa, shared across the whole tree), comparing TreeFlow variational inference (TreeFlow VI, teal) with BEAST 2 MCMC (BEAST 2, red). The solid curves show the estimated marginal posteriors and the shaded bands represent the Monte Carlo error of each estimate. For BEAST 2 the band is obtained by bootstrapping the MCMC samples; for TreeFlow VI it is estimated from the variability between four independent variational runs, with the curve showing the posterior pooled across those runs. The variational runs use the root full rank approximation; the corresponding comparison of all three approximation families is given in Supplementary Appendix Figure S3.}
    \label{fig:carnivoresmarginals}
\end{figure}

This problem could be approached by means of Bayesian model comparison; a lack of preference for a model allowing between-lineage variation of the ratio could indicate that the substitution model lacks the power to separate variation `signal' from saturation `noise'. Firstly, we construct a standard phylogenetic model with a HKY substitution model with a single transition-transversion ratio parameter (\textit{kappa}). We perform inference using ADVI, and also using MCMC as implemented in BEAST 2 (v2.7.7; \cite{bouckaert2019beast}). Since this model is of the form of a standard phylogenetic analysis, it could be fit using TreeFlow's command line interface. Figure \ref{fig:carnivoresmarginals} compares the marginal parameter estimates obtained from variational inference with the root full rank approximation in TreeFlow and MCMC in BEAST 2. The variational and MCMC estimates agree closely for every parameter of the substitution and site rate models, including all four base frequencies, and the two estimates of the parameter of interest, kappa, are almost indistinguishable. The discrepancies that remain are confined to quantities derived from the tree: the tree height is under-dispersed and the tree length over-dispersed. The root full rank approximation treats the node height ratios as independent, so it does not represent the correlations among node heights. The mode-seeking behaviour of the KL objective further favours narrow estimates. The corresponding comparison for the mean field and full rank families, which both show larger discrepancies, is given in Supplementary Appendix Figure S3.

We then altered this model to estimate a separate ratio for every branch of the tree, that is, one independent kappa parameter per branch; the implementation of this was simple and scalable as a result of TensorFlow's vectorization and broadcasting functionality. The model definition code for the base model and the small changes required to model variation in the kappa parameter across lineages are shown in Supplementary Appendix Figure S2. We compared the models using estimates of the marginal likelihood \cite{mackay2003information}. The \textit{marginal likelihood}, or \textit{evidence}, the integral of the likelihood of the data over model parameters, is typically analytically intractable and challenging to compute using MCMC methods \cite{xie2011improving}. The closed-form approximation to the posterior distribution provided by variational inference can be used as a proposal distribution for importance sampling \cite{burda2015importance}, a utility provided by TreeFlow. Concretely, one draws samples from the fitted approximation and weights each sample by the ratio of the model's joint density (the likelihood multiplied by the prior) to the density of the approximation. The average of these importance weights is an unbiased Monte Carlo estimate of the marginal likelihood. Because the importance weights account for the discrepancy between the variational approximation and the true posterior, this produces more accurate marginal likelihood estimates than the variational bound alone.

\begin{figure}
    \centering
    \begin{subfigure}[t]{0.4\linewidth}
        \includegraphics[width=\linewidth]{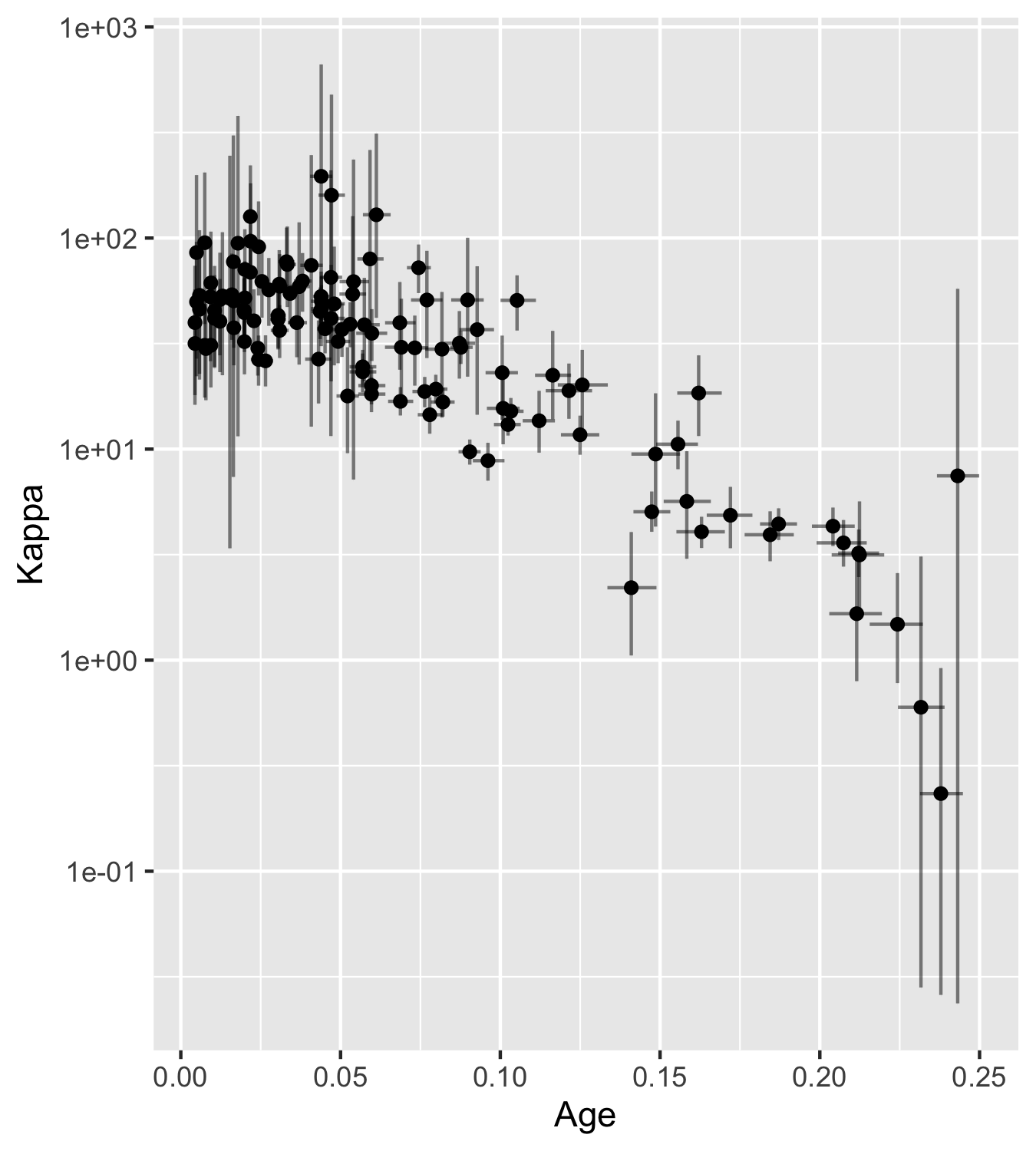}
        \caption{Estimated per-branch kappa against the age of each branch. Each point is the posterior mean of the kappa parameter for one branch, shown on a logarithmic scale; the vertical bars show its 95\% credible interval and the horizontal bars show the 95\% credible interval of the age. Age is the height of the branch's midpoint above the present, measured in expected substitutions per site (the tree is estimated in units of substitutions and is not calibrated to time).}
        \label{fig:carnivoreskappa}
    \end{subfigure}
    \begin{subfigure}[t]{0.4\linewidth}
        \includegraphics[width=\linewidth]{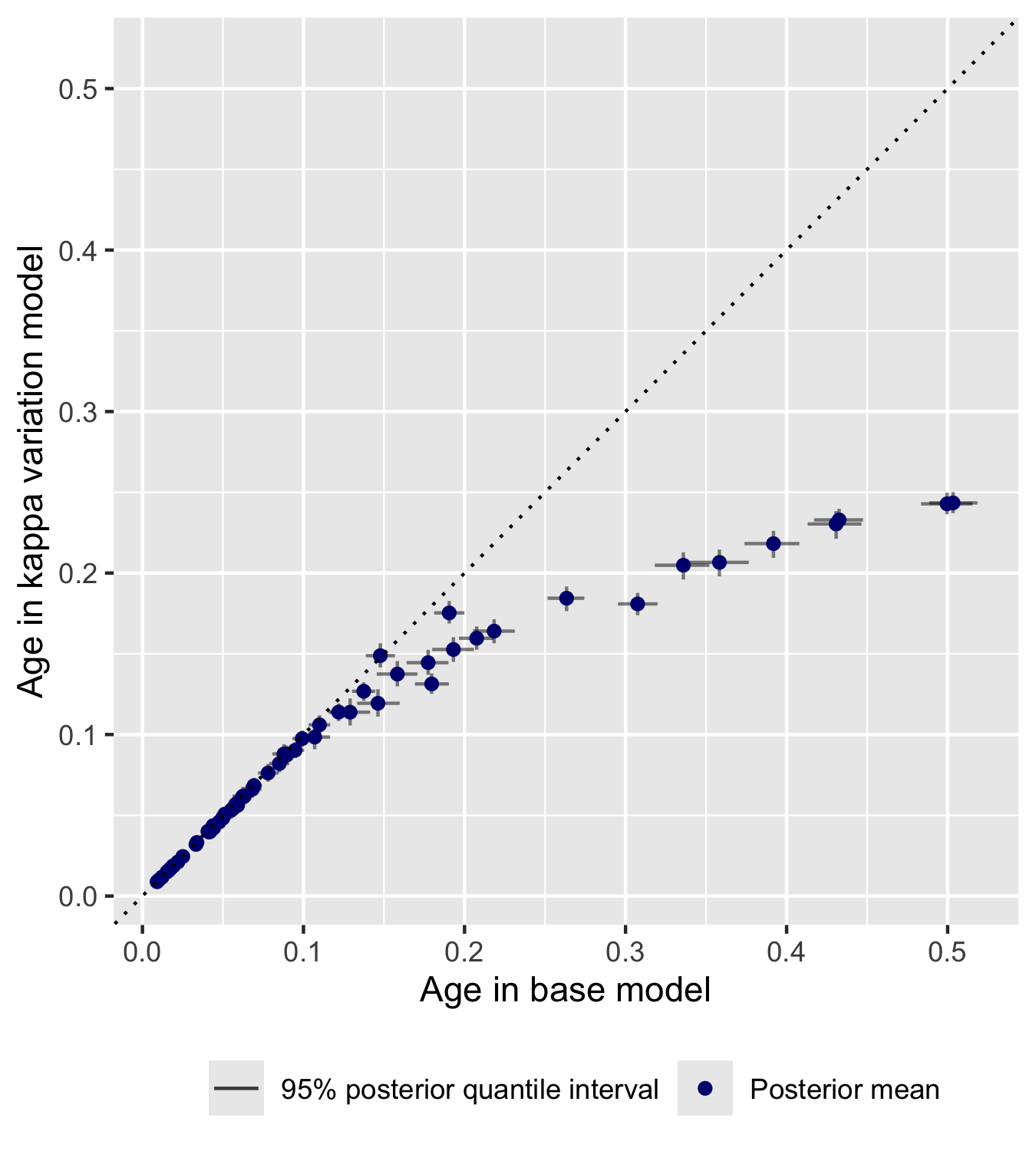}
        \caption{Internal node age estimates under the two models. Each point is one internal node, with its age under the base (single-kappa) model on the horizontal axis and its age under the per-branch kappa variation model on the vertical axis; points are posterior means, bars show the 95\% posterior quantile interval, and the dotted line is the identity. Points falling below the identity for older nodes show that the kappa variation model infers younger ages for deeper nodes.}
        \label{fig:carnivorestree}
    \end{subfigure}
    \caption{Parameter estimates for the carnivores per-branch transition/transversion ratio (kappa) variation model. In this model every branch is assigned its own independent kappa parameter, drawn from a shared log-normal prior sampled once per branch; kappa is not shared between branches and does not follow any model of change over time or at speciation events. Panel (a) shows how the per-branch kappa estimates vary with the age of the branch, and panel (b) shows the effect of allowing per-branch kappa variation on the estimated internal node ages.}
\end{figure}

The estimated per-lineage kappa parameters are shown in Figure \ref{fig:carnivoreskappa}. Kappa estimates decline with increasing age of the branch, which agrees with the results of the previous studies. The estimated log marginal likelihood for the base model was -193606.6 and for the model with kappa variation it was -192414.7, indicating that the model with transition-transversion ratio variation across lineages is preferred. This means that, under the other components of this model, the data supports variation in the ratio. This does not necessarily imply that transition-transversion ratio in the true generative process of the data varies in the same way, but could be a useful consideration in designing more sophisticated models of nucleotide substitution. The growth in uncertainty in kappa estimates deeper in the tree supports previous conclusions that nucleotide substitution models are unable to effectively estimate the number of substitutions on older branches \cite{duchene2015declining}. Figure \ref{fig:carnivorestree} shows that the kappa variation model shortens older branches, leading to a substantially reduced overall tree height estimate, with proportionally similar uncertainty in node height estimates. This is not a proper dating analysis because it does not account for uncertainty in the mutation rate and lacks time calibration. However, it is evident that the inclusion or exclusion of heterogeneity in the transition-transversion ratio affects the time estimates. This example demonstrates the power of TreeFlow's probabilistic modelling approach: extending the standard model to include per-lineage kappa variation required only a few lines of code changes (Supplementary Appendix Figure S2), and the variational inference and marginal likelihood estimation machinery could be applied to the new model without any additional implementation effort.

\begin{figure}
    \centering
    \begin{subfigure}[b]{\linewidth}
        \includegraphics[width=0.85\linewidth]{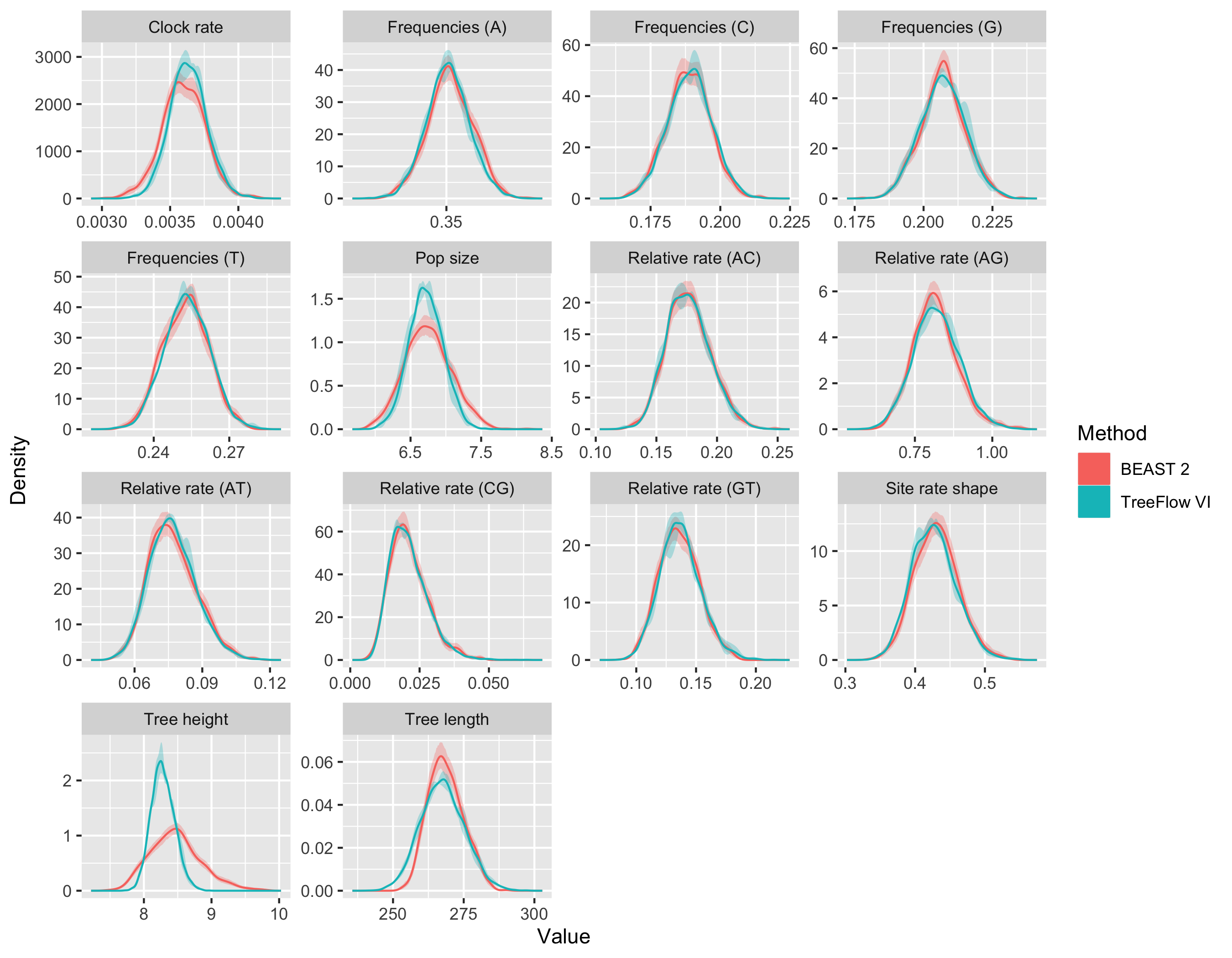}
        \caption{Marginal posterior distributions for the model parameters, comparing TreeFlow variational inference (TreeFlow VI, teal) with BEAST 2 MCMC (BEAST 2, red). Shaded bands are Monte Carlo error, obtained as in Figure \ref{fig:carnivoresmarginals} by bootstrapping for BEAST 2 and from four independent variational runs for TreeFlow VI, whose pooled posterior is the plotted curve. The variational runs use the root full rank approximation; Supplementary Appendix Figure S4 gives the corresponding comparison of all three approximation families.}
        \label{fig:flumarginals}
    \end{subfigure}

    \begin{subfigure}[b]{\linewidth}
        \includegraphics[width=0.85\linewidth]{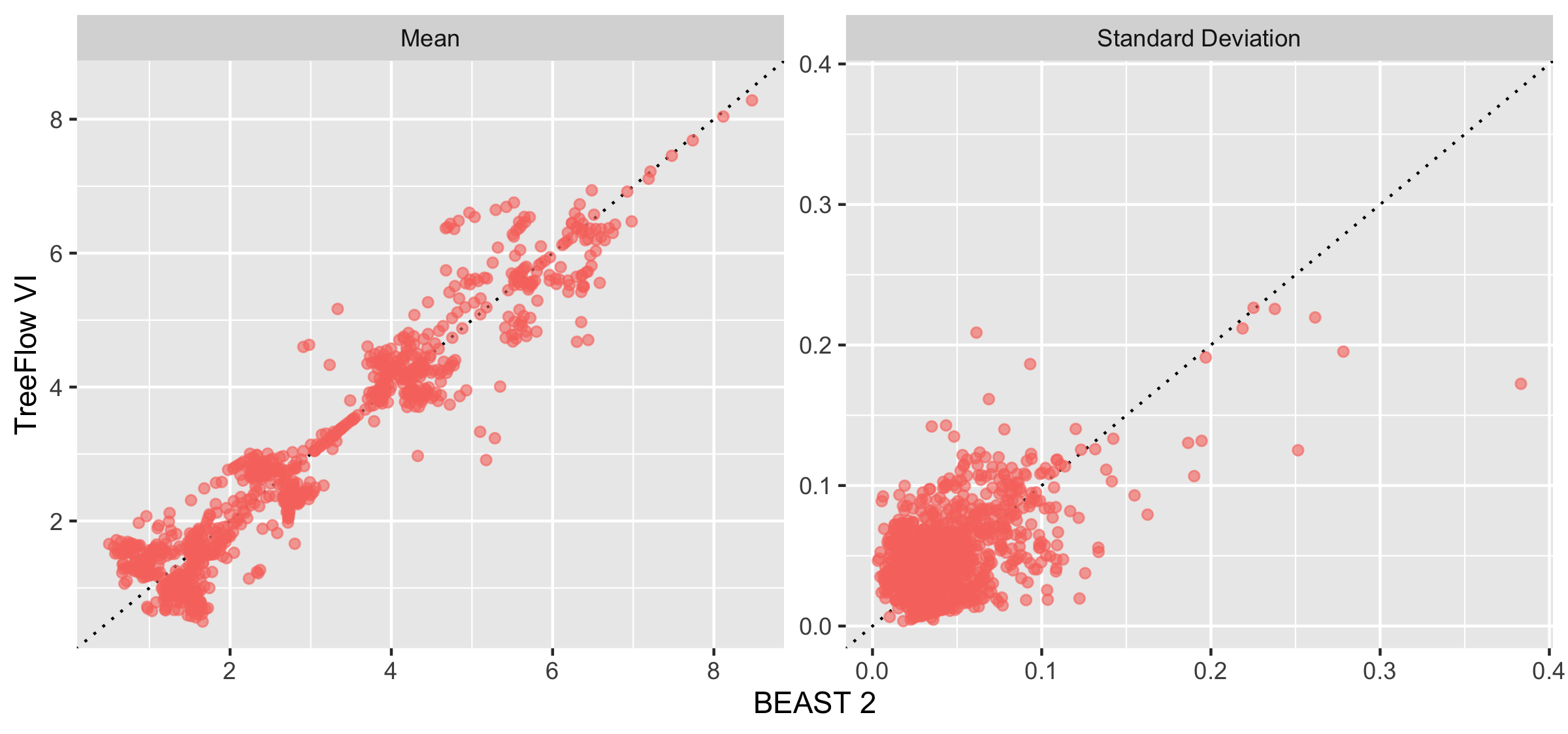}
        \caption{Per-node summaries of the internal node height posteriors, plotting the TreeFlow VI estimate against the BEAST 2 estimate for each node. The left panel compares the posterior means and the right panel the posterior standard deviations; the dotted line in each panel is the identity. Points above the identity in the right panel are node heights whose posterior uncertainty the variational approximation overestimates relative to BEAST 2, and points below it are ones it underestimates. The two occur together: the approximation is over-dispersed for the nodes BEAST 2 estimates most precisely, near the origin, and under-dispersed for the most uncertain ones.}
        \label{fig:flutree}
    \end{subfigure}

    \caption{Parameter and divergence-time estimates from the analysis of the influenza dataset, comparing TreeFlow variational inference with BEAST 2 MCMC.}
\end{figure}

\subsection{Scalable inference}

In the analysis of the 980-taxon influenza dataset, we demonstrate the scalability of variational inference. We performed variational inference using TreeFlow's command line interface with the root full rank approximation. The model used for inference included a coalescent prior with a constant population size on the tree, a strict molecular clock, a discretized Gamma model of site rate variation and a GTR substitution model. The YAML model definition code for this model, including parameter priors, is shown in Supplementary Appendix Figure S1. This parameterization of the GTR substitution model (here named \texttt{gtr\_rel}), with independent priors on five of the relative rates and one held fixed to 1, is used to allow comparison of parameter estimates with BEAST 2. It is also possible to use a six-rate GTR parameterization with a Dirichlet prior in TreeFlow such as used in other phylogenetics software \cite{huelsenbeck2001mrbayes, hohna2016revbayes}.

Figure \ref{fig:flumarginals} shows the marginal parameter estimates obtained from this dataset, compared to those obtained from a fixed-topology MCMC analysis using BEAST 2. The posterior distributions of the substitution and site rate model parameters, including all four base frequencies, are approximated with high fidelity. Similar to the carnivores dataset, the discrepancies are confined to the tree and to the parameters coupled to it: the tree height, coalescent effective population size and clock rate are under-dispersed, and the tree length is over-dispersed. The independent treatment of the node height ratios again discards the correlations among node heights, and the MCMC posteriors for the tree height and population size are additionally right-skewed, which the Normal-family approximation cannot reproduce.

Figure \ref{fig:flutree} compares the divergence time estimates. Under the root full rank approximation the root height is included in the covariance block, but the remaining node heights are represented by independent ratios transformed through the node height ratio transformation to respect the time constraints of the tree. The true posterior is almost certainly not of this form, so the approximation introduces error.

In general, the mean of the posterior distribution is well approximated, in particular for the oldest seven nodes of the tree. The posterior means of other divergence times are generally close but not identical to the true posterior. The error is more apparent in the posterior uncertainty; the standard deviation estimates produced by variational inference often differ substantially from the posterior estimated by BEAST 2. While the root full rank approximation seems effective for estimating the parameters of the phylogenetic model the estimates of divergence times appear less reliable. Variational approximations that better capture the form of the true posterior would improve the quality of these estimates.

BEAST 2 MCMC sampling was run for 30,000,000 iterations using the standard single-threaded MCMC routine with the BEAGLE library \cite{ayres2019beagle}. Convergence was checked using \textit{effective sample size} (ESS), which was computed for all parameters using ArviZ \cite{kumar2019arviz}. BEAST 2's built-in operator autotuning was used, and multiple preliminary runs were performed to manually adjust operator weights to improve ESSs. Generally, the minimum acceptable ESS is 100, though 200 is preferred \cite{drummond2015bayesian}. The analysis resulted in a minimum effective sample size of 103. The wall clock runtime for the BEAST 2 analysis was 6 hours and 21 minutes. Convergence for the TreeFlow-based variational inference analysis was assessed by monitoring the ELBO over the optimization routine and visually examining parameter traces. In practice, TreeFlow's command line interface reports an ELBO estimate at the end of optimization, and the full trace of the loss and the variational parameters over the optimization can be saved and plotted. Convergence is diagnosed by checking that the ELBO has reached a stable plateau and that the variational parameters have stopped drifting, and, if in doubt, by rerunning with a larger number of optimization steps. Step-by-step guidance on monitoring the convergence of variational inference in TreeFlow, including interpreting these traces, is provided in the online manual. The TreeFlow analysis converged in 22,000 iterations, which took 10 hours and 30 minutes; running the full 30,000 iterations took 14 hours and 20 minutes. Reaching convergence was therefore around 1 times faster than the BEAST 2 analysis was to reach the effective sample sizes reported above, and even the full optimization run, continued well past the point at which the ELBO had plateaued, took about a quarter of the MCMC runtime. This demonstrates that variational inference with TreeFlow is a viable option for large phylogenetic datasets, providing substantial speed improvements over MCMC despite the more expensive gradient computations required. TreeFlow's phylogenetic likelihood implementation performs well enough to be useful in real-world inference scenarios, and the entire analysis was specified using TreeFlow's YAML model definition format (Supplementary Appendix Figure S1) without any custom code.

\section{Benchmarks}
\label{benchmarks}

The phylogenetic likelihood is the computation that dominates the computational cost of model-based phylogenetic inference. We benchmark the performance of our TensorFlow-based likelihood implementation against specialized libraries. Since gradient computations are of equal importance to likelihood evaluations in modern automatic inference regimes, we also benchmark computation of derivatives of the phylogenetic likelihood, with respect to both the continuous elements of the tree and the parameters of the substitution model. A clear difference emerges in the implementation of derivatives; TreeFlow's are based on automatic differentiation while bespoke libraries need analytically-derived gradients. Therefore, we do not necessarily expect our implementation to be as fast as bespoke software, but it does not rely on analytical derivation of gradient expressions for every model and therefore automatically supports a wider range of models.

We benchmark both of the phylogenetic likelihood implementations described in the Tree traversals section: the portable, pure-TensorFlow TensorArray traversal, which we refer to as TreeFlow, and the compiled native C++ operation, which we refer to as TreeFlow (native).

We compare the performance of TreeFlow's likelihood implementation against BEAGLE \cite{ayres2019beagle}. BEAGLE is a library for high-performance computation of phylogenetic likelihoods. Recent versions of BEAGLE implement analytical gradients with respect to tree branch lengths \cite{ji2020gradients}. We use BEAGLE via the \texttt{bito} software package \cite{bito}, which provides a Python interface to BEAGLE's branch length gradients and also numerically computes derivatives with respect to substitution model parameters using a finite differences approximation. This hybrid approach has been shown to perform as well as fully analytical substitution model parameter gradients on large datasets \cite{fourment2023automatic}.

We also compare TreeFlow with a simple likelihood implementation \cite{phylojax} based on another automatic differentiation framework, JAX \cite{jax2018github}. JAX uses an eager execution model where execution of an operation immediately results in the computation being performed. JAX can additionally trace and compile a function ahead of execution with the XLA compiler (just-in-time, or JIT, compilation); we benchmark both the eager and JIT-compiled variants, denoted JAX and JAX (JIT). In contrast, TensorFlow's function mode, which is used by TreeFlow in the benchmarks, constructs a graph representation of the computation which is processed and can then be executed. As a result the phylogenetic likelihood in JAX can be implemented with native Python control flow even with a dynamic tree topology.

Benchmarks are performed on simulated data. Simulation allows the generation of a large number of data replicates with appropriate properties. Since we want to investigate the scaling of likelihood and gradient calculations with respect to the number of sequences, we simulate data for a range of sequence counts. Sequence counts are selected as increasing powers of 2 to better display asymptotic properties of the implementations. We simulate data with sequence counts ranging from 32 to 2048. Trees are simulated under a constant-size coalescent model for a given number of taxa, and then nucleotide sequences of length 1000 are simulated under a strict molecular clock and a Jukes-Cantor substitution model with no site rate variation. Tree and sequence simulation are performed using the sampling functionality implemented by TreeFlow's distributions.

We benchmark four distinct computations on these datasets. Firstly, for each replicate we compute the phylogenetic likelihood under a very simple model of sequence evolution. This uses a Jukes-Cantor model of nucleotide substitution and no model of site rate variation. Secondly, we calculate derivatives with respect to branch lengths under this simple model. Thirdly, we compute the likelihood under a more sophisticated substitution model, with a discretized Weibull distribution with four rate categories to model site rate variation and a GTR model of nucleotide substitution. This Weibull distribution is used for site rate variation in \texttt{bito} instead of the widely used Gamma distribution because its inverse cumulative distribution function (required for computing rates for the discretized categories) has a simpler form. Because this is a minor part of the computation, only performed once for each rate category per likelihood evaluation, TreeFlow's scaling should be unaffected if the standard Gamma distribution is used. Finally, we compute derivatives with respect to both branch lengths and the parameters of the substitution model (the six GTR relative substitution rates, four base nucleotide frequencies, and the Weibull site rate shape parameter).

All reported times are the cost of a \textit{single} evaluation of the computation in question. Each computation is timed individually over 100 repeated calls on one fixed input, and we take the minimum of those timings as the estimate for that dataset. This procedure is repeated on 10 independently simulated datasets per taxon count, and the figures and tables below report the mean across those datasets.

\begin{figure}
    \centering
    \includegraphics[width=\linewidth]{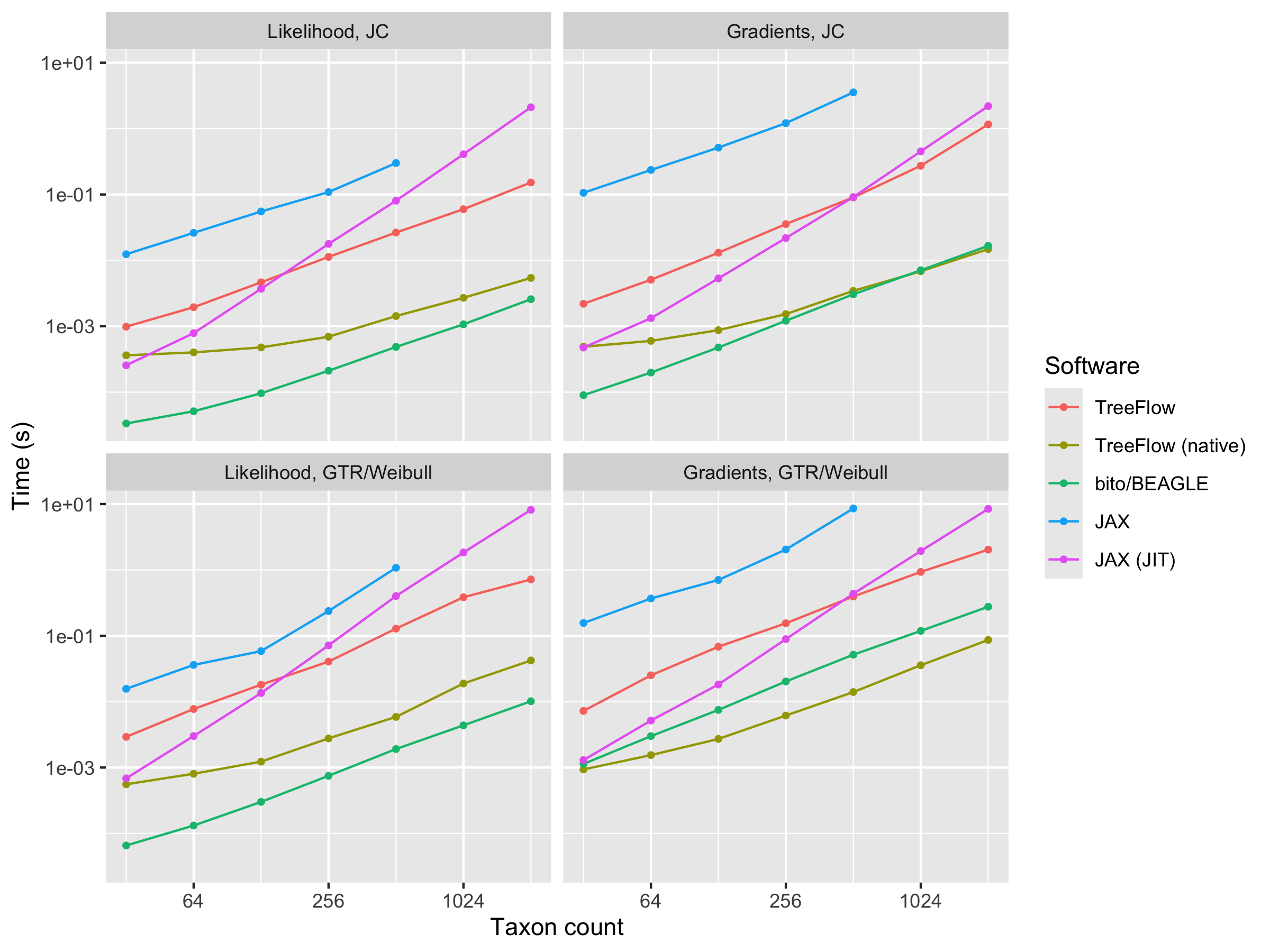}
    \caption{Times from the phylogenetic likelihood benchmark, on log-log axes. Each point is the time for a single evaluation of the computation on an alignment of 1000 sites, averaged over 10 simulated datasets. Each panel corresponds to one of the four benchmarked computations (likelihood and gradient evaluation, under the simple JC model and the more complex GTR/Weibull model), with the five implementations plotted as separate series: TreeFlow's portable TensorArray traversal (TreeFlow) and its native C++ operation (TreeFlow (native)), \texttt{bito}/BEAGLE, and the eager (JAX) and JIT-compiled (JAX (JIT)) JAX implementations. Eager JAX is only run up to 512 taxa as its runtime becomes prohibitive on larger trees.}
    \label{fig:benchmark}
\end{figure}

\begin{table}
    \centering
    \begin{tabular}{lllrr}
\toprule
 &  &  & Time per evaluation, s (512 taxa) & log-log slope \\
Model & Computation & Method &  &  \\
\midrule
\multirow[t]{10}{*}{JC} & \multirow[t]{5}{*}{Likelihood} & TreeFlow & 0.0264 & 1.22 \\
 &  & TreeFlow (native) & 0.00143 & 0.67 \\
 &  & bito/BEAGLE & 0.000485 & 1.07 \\
 &  & JAX & 0.300 & 1.12 \\
 &  & JAX (JIT) & 0.0804 & 2.20 \\
 & \multirow[t]{5}{*}{Gradients} & TreeFlow & 0.0902 & 1.48 \\
 &  & TreeFlow (native) & 0.00344 & 0.85 \\
 &  & bito/BEAGLE & 0.00305 & 1.27 \\
 &  & JAX & 3.54 & 1.25 \\
 &  & JAX (JIT) & 0.0913 & 2.05 \\
\multirow[t]{10}{*}{GTR/Weibull} & \multirow[t]{5}{*}{Likelihood} & TreeFlow & 0.129 & 1.36 \\
 &  & TreeFlow (native) & 0.00587 & 1.07 \\
 &  & bito/BEAGLE & 0.00191 & 1.24 \\
 &  & JAX & 1.08 & 1.49 \\
 &  & JAX (JIT) & 0.401 & 2.29 \\
 & \multirow[t]{5}{*}{Gradients} & TreeFlow & \bfseries 0.394 & \bfseries 1.34 \\
 &  & TreeFlow (native) & \bfseries 0.0140 & \bfseries 1.11 \\
 &  & bito/BEAGLE & \bfseries 0.0516 & \bfseries 1.33 \\
 &  & JAX & \bfseries 8.56 & \bfseries 1.40 \\
 &  & JAX (JIT) & \bfseries 0.435 & \bfseries 2.13 \\
\bottomrule
\end{tabular}

    \caption{Results of phylogenetic likelihood benchmark. Times are for a single evaluation of the computation on a 512-taxon tree with 1000 sites, in seconds, quoted to three significant figures and averaged over 10 simulated datasets; slopes are fitted to log-transformed times against log-transformed taxon counts across the full range of taxon counts. Times for gradients for the GTR/Weibull are highlighted as they are the most relevant computation for gradient-based inference on real data.}
    \label{tab:benchmarkfit}
\end{table}

Figure \ref{fig:benchmark} shows the results of the benchmarks with a log scale on both axes, and Table \ref{tab:benchmarkfit} reports representative runtimes together with the fitted log--log slopes. Considering first the phylogenetic likelihood, \texttt{bito}/BEAGLE is the fastest implementation, as expected for a highly optimized special-purpose library. TreeFlow's native operation is within a small constant factor of it: around three to four times slower for likelihood evaluation on the larger trees, with the gap widening to roughly an order of magnitude on the smallest trees, where fixed per-call overhead in the TensorFlow dispatch rather than the traversal itself dominates the runtime. It is in turn faster than TreeFlow's portable TensorArray implementation by a factor that grows from a few times on the smallest trees to one to two orders of magnitude on trees of a few hundred taxa and above. The TensorArray implementation is the slowest of the TreeFlow variants but, as discussed below, still scales sub-quadratically.

The picture changes for the gradient of the more complex GTR/Weibull model, which is the computation most relevant to gradient-based inference: for variational inference and Hamiltonian Monte Carlo the gradient is the primary quantity computed at each iteration. For this computation TreeFlow's native operation is the fastest of all implementations, faster even than \texttt{bito}/BEAGLE (Table \ref{tab:benchmarkfit}). The two are comparable on the smallest trees, but TreeFlow pulls ahead as the trees grow, reaching a factor of around three to four from a few hundred taxa upwards. The reason is the difference in how substitution-model parameter derivatives are obtained: \texttt{bito} computes them by finite differences \cite{fourment2023automatic}, performing additional likelihood evaluations per parameter, whereas TreeFlow obtains the entire gradient, with respect to branch lengths and all substitution-model parameters, in a single reverse-mode automatic-differentiation pass at close to the cost of one likelihood evaluation. This advantage grows with the number of parameters, so TreeFlow is an especially attractive choice for richly parameterized models (e.g. amino acid substitution models \cite{adachi1996model}) or models where the number of parameters grows with the number of sequences, as in the carnivore example above. Even TreeFlow's portable TensorArray implementation, though slower in absolute terms, benefits from the same advantage on the many-parameter gradient.

The JAX-based implementations are the slowest overall. In eager mode JAX is several times to nearly two orders of magnitude slower than TreeFlow's TensorArray implementation, with the largest gaps on the gradient computations, and its runtime becomes prohibitive on large trees, so we cap the eager-JAX benchmark at 512 taxa. JIT-compiling the JAX computation with XLA yields a large speed-up at small tree sizes, so that at a few hundred taxa it is comparable to TreeFlow's TensorArray implementation, but its runtime scales quadratically, with a log--log slope close to two (Table \ref{tab:benchmarkfit}), so that on the largest trees it becomes the slowest implementation of all. This behaviour illustrates the quadratic-complexity pitfall described in the Tree traversals section: a naive automatic-differentiation implementation of a tree traversal incurs a cost that grows with the square of the number of taxa. Just-in-time compilation reduces the constant factor of the computation, which is what gives the JIT-compiled variant its advantage at small taxon counts, but it does not change the underlying algorithmic complexity, so the quadratic scaling eventually dominates and the small-taxa advantage is lost.

Table \ref{tab:benchmarkfit} shows the coefficients obtained from fitting a linear model to the benchmark times where the predictor is the log-transformed number of sequences and the target is the log-transformed runtime. The slope parameter estimates from these fits are a rough empirical estimate of the polynomial degree of the computational scaling. The slopes for all of the TreeFlow and \texttt{bito}/BEAGLE computations are well below 2, indicating sub-quadratic scaling with the size of the data; the native operation has the smallest slopes of any implementation. By contrast the JIT-compiled JAX implementation has slopes close to 2 across all four computations, consistent with the super-linear behaviour noted above. For the JC benchmark, the only gradients computed are with respect to branch lengths, so the whole computation for \texttt{bito}/BEAGLE is the analytical linear-time branch gradient. In this case TreeFlow's TensorArray slope is steeper than that of \texttt{bito}/BEAGLE (Table \ref{tab:benchmarkfit}). Both are nonetheless far below the slopes above two returned by the JIT-compiled JAX implementation, which is what a quadratic traversal produces on the same data, and TreeFlow's native operation has a shallower slope than \texttt{bito}/BEAGLE's analytical gradient. The TensorArray implementation therefore achieves sub-quadratic scaling for branch gradients through automatic differentiation, in the same regime as the hand-derived analytical gradient, even though it is a constant factor slower.

Taken together, these results show that TreeFlow's native operation makes automatic-differentiation-based phylogenetics competitive with a specialized native library, and faster than it for the most inference-relevant computation, the gradient of a many-parameter model. The TensorArray traversal implementation provides an alternative with similar sub-quadratic scaling without requiring a hand-derived gradient for new models.

\section{Discussion}

The probabilistic modelling approach enabled by the TreeFlow Python API in conjunction with TensorFlow Probability has similar goals to other phylogenetic probabilistic modelling software. Two notable examples, RevBayes \cite{hohna2016revbayes} and Blang \cite{bouchard2022blang} provide a generic model definition and inference engine which also support phylogenetic models and tree inference. Some notable differences exist between TreeFlow and these which may prove advantageous for certain users. Firstly, RevBayes requires the user to manually specify an MCMC operator scheme, including parameters and weights. This may require substantial user time and expertise to achieve correct and efficient inference. Blang and TreeFlow, on the other hand, are focused on automatic inference methods that require less user input. Secondly, TreeFlow and its Python model specification API are implemented in a widely used computational framework (TensorFlow) and programming language (Python). As a result it can potentially leverage other computational tools implemented in TensorFlow, including less common distributions, machine learning models, differential equation solvers and alternative inference schemes. The automatic differentiation capabilities TensorFlow affords mean TreeFlow can interface with the large and growing collection of models that rely on gradient-based inference. On the other hand, the choice of computational framework incurs a performance cost as evidenced by the results of benchmarking. Blang models can make use of external Java code, but the ecosystem of related software in Java is less developed, and RevBayes does not provide any functionality for interfacing with software that is not implemented in the Rev language.

The combination of flexibility and scalability of libraries like TreeFlow has the potential to be useful for rapid analysis of large modern genetic datasets, such as those assembled for tracking infectious diseases \cite{hadfield2018nextstrain}. For this purpose, TreeFlow's true power would be unlocked with the implementation of improved modelling and inference tools for evolutionary rates and population dynamics.

For principled dating analyses, uncertainty in evolutionary rate parameters must be considered. Posterior distributions involving these parameters may have significant correlation structure which would be ignored by variational inference with approximations that assume independence. These parameters typically receive special treatment in MCMC sampling routines \cite{drummond2006relaxed, zhang2020improving}. As observed in our analysis on the influenza dataset, the independence between ratios of internal node heights assumed by the root full rank approximation fails to accurately capture the posterior distribution on all internal divergence time estimates. Implementing structured posterior approximations for these models could substantially improve the reliability of parameter estimates obtained using variational inference while scaling to extremely large datasets. For example, the known correlation between the clock rate and tree height, or correlations between adjacent nodes in the tree, could be captured by adding additional covariance structure to the variational family without incurring the cost of a full-rank covariance matrix. Correlations that are not linear in the unconstrained coordinates, and distribution shapes outside the Normal family could be addressed by applying normalizing flows \cite{rezende2015variational} to model more complex dependencies.

Additionally, implementing flexible models for population dynamics, such as nonparametric coalescent models \cite{drummond2005bayesian, minin2008smooth, gill2013improving}, could provide valuable insights from large datasets. These models would result in complex posterior distributions, which would need to be accounted for in a variational inference scheme. If these coalescent models could be made to work effectively in TreeFlow, the probabilistic graphical modelling paradigm would allow for rapid computational experimentation with novel models of time-varying population parameters.

Finally, the most significant functionality for phylogenetic inference missing from TreeFlow is inference of the tree topology. TreeFlow's tree representation could allow for the topology to become an estimated parameter, such as in existing work on variational Bayesian phylogenetic inference \cite{zhang2018variational}. Efficiently implementing the computations required for these algorithms in the TensorFlow framework would be a major computational challenge, but would be supported by TreeFlow's representation of the tree topology as a dynamic variable inside the computational graph. A promising direction would be to combine TreeFlow's existing model components and gradient computations with a variational distribution over tree topologies, such as the subsplit Bayesian network approach \cite{zhang2018generalizing}, enabling joint inference of topology and continuous parameters.

\section{Conclusion}

TreeFlow is a software library that provides tools for statistical modelling and inference on phylogenetic problems. Probabilistic graphical modelling provides flexible model definition involving phylogenetic trees and molecular sequences, and automatic differentiation enables the application of modern scalable inference algorithms with a fixed tree topology. TreeFlow's automatic differentiation-based implementations of core gradient computations provide reasonable performance compared to specialized phylogenetic libraries. These building blocks have the potential to be used in valuable phylogenetic analyses of large and complex genetic datasets, and with the next generation of inference methods that can accelerate the inference of tree topologies using gradient information.

\section{Availability}

TreeFlow is an open-source Python package. Instructions for installing TreeFlow, and examples of using it as a library and command line application can be found at \url{https://github.com/christiaanjs/treeflow}. The manual for TreeFlow can be found at \url{https://treeflow.readthedocs.io}.

\section{Data availability}

Datasets, model definitions, starting values, and BEAST 2 XML files used in the biological examples are available at \url{https://github.com/christiaanjs/treeflow-paper/releases/tag/v0.0.1} or \url{https://doi.org/10.5061/dryad.v6wwpzh0p}.

\section{Conflicts of Interest}

The authors declare that there are no conflicts of interest.

\bibliography{treeflow}

\end{document}